
\magnification=\magstep1
\hsize 32 pc
\vsize 42 pc
\baselineskip = 24 true pt
\def\cl{\centerline}
\cl {\bf Exact Solution of a N-body Problem in One Dmension}
\vskip .5 true cm
\cl {\bf Avinash Khare$^*$}
\vskip .2 true cm
\cl {Institute of Physics, Sachivalaya Marg,}
\cl {Bhubaneswar-751 005, India}
\vskip .5 true cm
\noindent {\bf Abstract}

\vskip .3 true cm
Complete energy spectrum is obtained for the quantum mechanical
problem of N one dimensional equal mass particles
interacting via potential
$$V(x_1,x_2,...,x_N) = g\sum^N_{i < j}{1\over (x_i-x_j)^2} -
{\alpha\over \sqrt{\sum_{i < j} (x_i-x_j)^2}}$$
Further, it is shown that scattering configuration, characterized by
initial momenta $p_i (i=1,2,...,N)$ goes over into a final
configuration characterized uniquely by the final momenta $p'_i$ with
$p'_i=p_{N+1-i}$.
\vfill
* e-mail address : khare@iopb.ernet.in
\eject
In recent years, the Calogero - Sutherland (CS) type of N-body
problems in one dimension have received considerable attention in the
literature [1,2,3,4]. It is believed that the CS model with inverse
square interaction provides an example of an ideal gas in one
dimension with fractional statistics [5]. Besides, these models are
related to (1+1) - dimensional conformal field theory, random matrices
as well as host of other things [6]. Inspired by these successes, it
is of considerable interest to discover new exactly solvable N-body problems.

The purpose of this note is to present one such example. In
particular I show that the N-body problem with equal mass in
1-dimension characterized by $(\hbar = 2m = 1,\ g > - 1/2,\ \alpha > 0)$
$$H = - \sum^N_{i=1} {\partial^2\over\partial x_i^2} +\sum^N_{i<j}
{g\over (x_i-x_j)^2} - {\alpha\over \sqrt{\sum_{i < j}
(x_i-x_j)^2}}\eqno {(1)}$$
is exactly solvable. The interesting point about this model is that
unlike most other exactly solvable models,
it has both bound state and scattering solutions. In particular I
show that the complete bound state spectrum ( in the center-of-mass
frame) is given by the formula
$$E_{n+k} = - {\alpha^2\over 4N[n+k+b + {1\over 2}]^2} , \ \ n,k =
0,1,2,\eqno{(2)}$$
where
$$b =  {N (N-1)\over 2} a +  {N (N+1)\over 4} - {3\over 2} ; \ \ a =
{1\over 2} \sqrt{1+2g}\eqno{(3)}$$

For positive energy one has only scattering states. I show that a
scattering configuration, characterized by initial momenta $p_i
(i=1,2,...,N)$ goes over into final configuration characterized
uniquely by the final momenta $p'_i$ with
$$p'_i = p_{N+1-i}\eqno{(4)}$$
However, unlike the pure inverse square scattering case $(\alpha =
0)$, in our case the phase shift is energy dependent. Thus, as in
other integrable cases, in our case too the scattering problem reduces
to a sequence of 2-body processes.

Finally, a la Sutherland [3], I also solve a slightly diferent variant of
the Hamiltonian  (1) with $-\alpha/\sqrt{\sum_{i<j}(x_i-x_j)^2}$
being replaced by an external potential $-\alpha/\sqrt{\sum_ix_i^2}$
and obtain exact expressions for the ground state energy eigenvalues
and eigenfunctions.

Consider the Hamiltonian as given by eq. (1). We need to solve the
eigenvalue equation
$$H\psi = E\psi\eqno{(5)}$$
where $\psi$ is a translation invariant eigenfunction. Note that
our Hamiltonian is very similar to the classic Calogero Hamiltonian
(see eq. (2.1) of his paper [2]) except that whereas he has a pairwise
quadratic potntial, we have a "N-body" potential
$-\alpha \over \sqrt{\sum_{i<j}(x_i-x_j)^2}$. However, we shall see that
many of the key steps are very similar in the two cases and hence we
avoid giving most steps which are already contained there [2].
Without any loss of generality, a la [2] we also restrict our attention
to the sector
of the configuration space corresponding to a definite ordering of
particles, say
$$x_i \geq x_{i+1}, \ \ i = 1,2,...,N-1.\eqno{(6)}$$
A la [2] it is clear that the normalizable solutions of eq. (5) (with H
being given by eq. (1)) can be cast in the form
$$\psi (x) = Z^{a+1/2} \phi (r) P_k (x)\eqno{(7)}$$
where a is defined in eq.(3) while Z and r are given by
$$ Z = \Pi^N_{i <j} (x_i-x_j) ,\ \ r^2 = {1\over N}\sum^N_{i <j}
(x_i-x_j)^2,\eqno{(8)}$$
and
$P_k(x)$ is a homogeneous polynomial of degree k in the particle
coordinates and satisfies the generalized Laplace equation i.e.
$$\bigg [ \sum^N_{i=1}{\partial^2\over\partial x^2_i} + 2 (a+{1\over
2})\sum^N_{i<j} {1\over (x_i-x_j)} ({\partial\over \partial x_i} -
{\partial\over \partial x_j})\bigg ] P_k(x)=0\eqno{(9)}$$
As discussed in detail in [2], the polynomials $P_k(x)$ are completely
symmetrical under the exchange of any two coordinates. On inserting
the ansatz (7) into the Schr\"odinger eq. (5) (with H given by eq. (1))
and using eq. (9) and following the procedure of ref [2], we find that
$\phi (r)$ satisfies the equation
$$-\bigg [ \phi^"(r)+\{ 2k+2b+1\}{1\over r}\phi' (r) \bigg ] -
({\alpha\over \sqrt N r} +E)\phi (r) = 0\eqno{(10)}$$
where prime denotes differentiation with respect to the argument.
The normalizable solutions of this equation are
$$\phi_{n,k} (r) = exp (-\sqrt{\mid E\mid} r) L_n^{2b+2k}(2\sqrt{\mid
E\mid} r).\eqno{(11)}$$
while the corresponding energies are as given by eq. (2). Here
$L^{\alpha}_n (r)$ is a Laguerre polynomial. Notice that in the expression (2)
for the energy, n and k always come in the combination n+k (unlike in the
Calogero case [2] where it comes in the combination 2n+k).

In the special case of N = 3, we can check our expressions for $E_n$
and $\psi_{n,k}$ with the exact expressions obtained by entirely
different method (see eqs. (40) to (43) of [7] ).
On comparing the two we find (note coupling constant in [7] is
$\sqrt 3 \alpha$ rather than $\alpha$) that the two expressions agree
provided k = 3l and $P_k(x) \propto r^{3l} C_l^{a+1/2} (cos 3\phi)$
where $C^a_i$ is a Gegenbauer polynomial.

Let us now consider the positive energy spectrum of the Hamiltonian
(1). It is of course purely continuous spectrum. Following the
treatment given above and as in [2] (Sec. 4) it is clear that the
complete set of stationary eigenfunctions of the problem (in the
center-of-mass frame) is
$$\psi_{pk} = Z^{a+1/2}\phi_p (r) P_k(x), \ k = 0,1,2,...; \ \ p\geq 0
\eqno{(12)}$$
where p is connected to the energy eigenvalue by $E=p^2 \geq 0$ (note
that we have chosen $\hbar = 2m = 1$) while $\phi_p(r)$ satisfies
eq. (10). It is easily shown that for $E\geq 0,$ the solution of eq. (10)
is given by
$$\phi_p (r) = e^{ipr} F (k+b+{1\over 2}-{i\alpha\over {2p \sqrt N}}, 2k+2b+1;
-2ipr)\eqno{(13)}$$
One can now run through the arguments of [2] (Sec. 4) and show that
if the stationary eigenfunction describing, in the center-of-mass
frame, the scattering situation is characterized by the form
$$\psi_{in}\sim C (exp [ i \sum^N_{i=1}p_i x_i]\eqno{(14)}$$
with (note $x_i \geq x_{i+1}, \ \  i = 1,2,...,N-1)$
$$p_i\leq p_{i+1}, \ \ p^2 = \sum^N_{i=1} p^2_i, \ \ \sum^N_{i=1}p_i
=0\eqno{(15)}$$
then $\psi_{out}$ is given by
$$\psi_{out}\sim C e^{2i\eta_p-ib\pi} exp [ i \sum^N_{i=1} p_{N+1-i}
x_i]\eqno{(16)}$$
where
$$e^{2i\eta_p} = {\Gamma (k+b+{1\over 2}-{i\alpha\over {2p \sqrt N}})
\over \Gamma (k+b+{1\over 2}+{i\alpha\over {2p \sqrt N }})}\eqno{(17)}$$
 Thus we have the remarkable result that even in the presence of the
potential\hfil\break $-{\alpha \over \sqrt{\sum_{i<j}(x_i -x_j)^2}}$,  the
N-particle
scattering problem reduces to a sequence of 2-body processes as
characterized by eq. (4) but now one has an energy dependent phase shift.
Note that all the results about scattering are also valid in case $\alpha$
is negative but now the spectrum is purely continuous and there are no
bound states.

 Finally, let us discuss the ``Sutherland variant'' [3] of the
Hamiltonian (1). Consider
$$H = - \sum^N_{i=1} {\partial^2\over\partial x^2_i} +\sum_{i <j}
{g\over (x_i-x_j)^2} - {\alpha\over \sqrt{\sum_{i} x^2_i}}\eqno{(18)}$$
i.e. the N-body potential is now an external potential. Note
that for the Calogero case, Sutherland [3]  was able to obtain an exact
expression for the ground state energy and eigenfunction $\psi$ and find a
remarkable connection of $\psi^2$ with the
joint probability density function for the eigenvalues of matrices
from a Gaussian ensembles in case $\beta = 2\lambda$ = 1, 2 or 4. Using
these connections he was the able
to compute [3] the one particle density and the pair correlation function.

Following Sutherland, let us consider the Schr\"odinger equation
$H\psi = E\psi$ with H as given by eq. (18). Further, let us write
the wavefunction $\psi$ as $\psi = \phi\Phi$ with
$$\phi = \Pi_{i<j} \mid x_i-x_j\mid^{\lambda} ,\ \ \lambda = {1\over
2}+a\eqno{(19)}$$
On using this ansatz in the Schr\"odinger equation with H as given by
eq.(18) we find that $\Phi$ must satisfy
$$-\sum^N_{i=1} {\partial^2\Phi\over\partial
x^2_i}-2\lambda\sum_{i<j} {1\over (x_i-x_j)}({\partial\over \partial
x_i}-{\partial\over \partial x_j})\Phi-{e^2\over \sqrt{\sum_ix^2_i}}\Phi
= E \Phi\eqno{(20)}$$
It is easily verified that
$$\Phi = exp (-\sqrt{\mid E\mid}\sqrt{\sum_ix^2_i})\eqno{(21)}$$
is a solution to eq. (20) with the energy
$$ E = - {e^4\over [(N-1)(1+\lambda N)]^2}\eqno{(22)}$$
Clearly, for each ordering of particles, $\psi$ is nodeless and
hence it is the solution for the ground state. If we rewrite
$\psi$ in terms of the variables
$$y_i = {\sqrt{\mid E\mid }\over\sqrt{\lambda}} x_i\eqno{(23)}$$
then one finds that
$$\psi^2 = C(exp (-\beta\sqrt{\sum_{i} y^2_i}) \Pi_{i < j} \mid
y_i-y_j\mid^{\beta}\eqno{(24)}$$
where C is the normalization constant. A la original Sutherland case [3], where
$\psi^2$ was identical to the joint probability density function, it
would indeed be remarkable if our $\psi^2$ as given by eq. (24)
for atleast $\beta = 1, 2, 4$ can be
mapped on to some known solvable problem and using these results if
one could obtain the one particle density and the pair correlation function
for our case.

This work raises several issues which need to be looked into. I hope to
address some of these issues in the near future.

\vfill
\eject
\noindent {\bf References}

\item {[1]} F. Calogero, Jour. Math. Phys. {\bf 10} (1969), 2191, 2197.
\item {[2]} F. Calogero, Jour. Math. Phys. {\bf 12} (1971) 419.
\item {[3]} B. Sutherland, Jour. Math. Phys. {\bf 12} (1971) 246.
\item {[4]} For excellent review of this field see M.A. Olshanetsky
and A.M. Perelomov, Phys. Rep.{\bf 71} (1981) 313; {\bf 94} (1983) 313.
\item {[5]} F.D.M. Haldane, Phys. Rev. Lett. {\bf 67} (1991) 937;
Y.-S. Wu, ibid {\bf 73} (1994) 922, M.V.N. Murthy and R. Shankar, ibid
{\bf 73} (1994) 3331.
\item {[6]} See for example the flow chart with various connections
in B.D. Simons, P.A. Lee and B.L. Altshuler, Phys. Rev. Lett. {\bf 72}
(1994) 64.
\item{[7]} A. Khare and R.K. Bhaduri, J. Phys. {\bf A 27} (1994) 2213.
\item{[8]} See for example, M.L. Mehta, Random Matrices (Revised Edition),
Academic Press, N.Y. (1990).
\vfill
\eject
\end